\documentclass[reprint,prb,aps,showpacs,twocolumn]{revtex4-1}
\usepackage{epsfig,color}
\usepackage{graphicx}
\usepackage{amsmath}
\usepackage{amssymb}

\begin{document}
\title{Signatures of spin-orbit coupling in scanning gate conductance images\\ of electron flow from quantum point contacts}

\author{M. P. Nowak}
\affiliation{AGH University of Science and Technology, Faculty of Physics and Applied Computer Science,\\
al. Mickiewicza 30, 30-059 Krak\'ow, Poland}
\author{K. Kolasi\'nski}
\affiliation{AGH University of Science and Technology, Faculty of Physics and Applied Computer Science,\\
al. Mickiewicza 30, 30-059 Krak\'ow, Poland}
\author{B. Szafran}
\affiliation{AGH University of Science and Technology, Faculty of Physics and Applied Computer Science,\\
al. Mickiewicza 30, 30-059 Krak\'ow, Poland}

\date{\today}

\begin{abstract}
Electron flow through a quantum point contact in presence of spin-orbit coupling is investigated theoretically in
the context of the scanning gate microscopy (SGM) conductance mapping.
Although in the absence of the floating gate the spin-orbit coupling does not significantly alter the conductance,
we find that the angular dependence of the SGM images of the electron flow at the conductance plateaux is substantially altered as the spin-orbit interaction mixes the orbital modes that enter the quantum point contact.
The radial interference fringes that are obtained in the SGM maps at conductance steps are essentially preserved by the spin-orbit interaction as backscattering by the tip preserves the electron spin although the effects of the mode mixing are visible.
\end{abstract}


\maketitle
\section{Introduction}

Scanning gate microscopy (SGM) has become a widely used technique that allows for mapping the current flow and charge densities in nanoscopic structures.
The perturbation induced by the floating gate was used to map scarred wave functions in quantum billiards,\cite{crook,burke} local density of states in quantum rings,\cite{martins,pala2008,pala2009} magnetic focusing of electrons\cite{aidala} or for  demonstration of a mesoscopic analogue to the Braess paradox.\cite{pala2012}
Mapping of conductance of quantum point contacts (QPCs) allowed for observation of spatial maps of the coherent electron flow \cite{topinka2000,leroy2002,qpc0,qpc5,qpc6,qpc7,qpc8} and signatures
of interference involving the presence of the tip.\cite{abbout,kozikov}
Moreover it has been found that the electron current after leaving the QPC propagates in narrow branches\cite{topinka2001,jura2007} and that checkerboard interference patterns\cite{leroy2005} are observed in the maps of the electron flow.\cite{jura} Recent experiments measured nonequilibrium transport phenomena in QPCs\cite{jura2010} and demonstrated the control of the edge channel trajectories in the quantum Hall regime.\cite{paradiso}
Recent theoretical studies on SGM in QPC systems delivered a perturbative description\cite{jalabert} of the transport and a temperature-induced amplification of the interference fringes.\cite{abbout}

There is a growing interest in spin phenomena in two-dimensional electron gas (2DEG). A particular attention is paid to spin-orbit (SO) interaction that results in an effective magnetic field for propagating electrons and allows for control of the electron spin by the electric fields.\cite{nowack,perge} The majority of SGM experiments probing electron flow from QPCs focus on structures based on GaAs where the SO interaction is usually weak. On the other hand materials such as InGaAs provide much stronger SO coupling.\cite{park} QPCs in these structures can be used to generate spin-polarized currents in the absence of external magnetic field.\cite{meto,sablikov,kim,nowak2013} Nevertheless, the number of SGM studies of InGaAs based QPCs is still limited.\cite{aoki} The present work describes the SO coupling effects on SGM imaging of electron flow from InGaAs QPCs in conditions of strong SO coupling. Previously, the impact of SO interaction on SGM maps of electron flow has been studied in the context of spin-dependent magnetic focusing.\cite{reynoso} We find that the SO interaction barely modifies the conductance of the unperturbed system. Nevertheless, we demonstrate  that the SO coupling  leaves a distinct signature on the QPC conductance response to the perturbation introduced by the tip.

In the present work we focus on two types of QPC work points. The first one is set by the conductance plateaux $G=1,2,3$ in units of $G_0=[2e^2/h]$ where the experiments detect the angular patterns of the electron flow from the QPC.\cite{topinka2000}  We find that the angular patterns are modified by the SO interaction in such a way that the double and triple branches are smeared out due to mixing of the orbital modes of the propagating electron by the SO coupling. A similar effect is observed for the QPC tuned to the conductance steps. We find that in the latter case that the radial interference fringes that are resolved by the SGM \cite{jura} preserve their oscillation pattern in contrast to the results obtained for Zeeman effect present where a beating pattern is observed.\cite{klesh}

\section{Theory}

\subsection{Model}
\begin{figure}[ht!]
\epsfxsize=85mm
                \epsfbox[40 302 557 550] {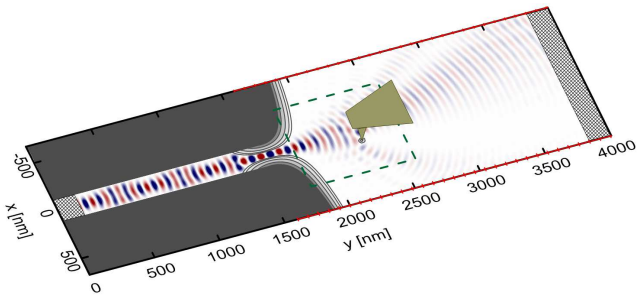}
                 \caption{(color online) Sketch of the system under consideration. The grey area depicts the confinement potential of the input channel. The contours
                  show the potential of a QPC for $\hbar \omega = 3$ meV for which we obtain the first plateau of conductance $G = 2 e^2/h$.  The real part of the
                  spin-up component of the scattering wave function for the electron incident from the channel is presented with the blue-red color map.  The tip is located at $x=100$ nm and $y=2300$ nm. The crossed parts at the input and output channels denote the regions used for resolution of the transmitted
                  and reflected plane waves. The red lines at the edges depict the regions where transparent boundary conditions are introduced. }
 \label{potbmp}
\end{figure}

We consider a two-dimensional system described by the Hamiltonian
\begin{equation}
H=\left[\frac{\hbar^2k_x^2}{2m^*}+\frac{\hbar^2k_y^2}{2m^*}+V(x,y)\right]\textbf{1} + \alpha(\sigma_x k_y -\sigma_y k_x),
\label{ham}
\end{equation}
where the latter term corresponds to SO coupling of Rashba type \cite{rashba} with the strength controlled by the parameter $\alpha$ and $\textbf{1}$ is identity matrix.
We consider the system which is schematically shown in
Fig. \ref{potbmp}, with a channel opening to an infinite space through a restriction introduced by the QPC.
The total potential in the system $V(x,y)$ is taken in the following form,
\begin{equation}
\begin{split}
V(x,y) = &V_{\mathrm{ch}}(x,y)+\frac{V_{\mathrm{tip}}}{(x-x_{\mathrm{tip}})^2/d^2+(y-y_{\mathrm{tip}})^2/d^2+1}\\
+&\exp\left\{ -\left[(y-Y_{\mathrm{qpc}})/(\sqrt{2}w)\right]^2 \right\}m^*\omega^2x^2/2,
\label{pot}
\end{split}
\end{equation}
where the first term describes the input channel
of width $W$
\begin{equation}
V_{\mathrm{ch}}(x,y)=\left\{
\begin{array}{ll}
      0 & |x|\leq W \\
      500\;\mathrm{meV} & |x| > W \mathrm{and}  < 1600\;\mathrm{nm}\\
\end{array}
\right.
\label{vch}
\end{equation}

The grey area in Fig. 1 shows the confinement potential of the input channel of Eq. (\ref{vch}).
The second term in Eq. (\ref{pot}) accounts for the effective potential of the tip localized above point $(x_\mathrm{tip},y_\mathrm{tip})$.
This potential is a result of interaction of the Coulomb charge at the tip and the 2DEG, which has a form
close to a Lorentzian.\cite{szafran2011} We adopt $d=20$ nm, $V_{\mathrm{tip}}=7$ meV as the parameters of the scanning probe, for
 which the conductance response that we obtain is comparable to the experimental values of Ref. \onlinecite{jura}.
The third term in Eq. (\ref{pot}) describes a smooth potential of the QPC with the center at $Y_\mathrm{qpc}$,\cite{nowak2013}
with $w$ responsible for the length of the constriction (we take $w=100$ nm unless stated otherwise).
We consider that the QPC is parabolic in the $x$ direction and described by the energy $\hbar\omega$.

\subsection{Calculation of the conductance}
We take $E_F=3$ meV and for most of the calculations we consider the input channel width of $W=240$ nm. In these conditions there are $N=8$ subbands at the Fermi level in the input channel including the spin degree of freedom -- see the dispersion relation in the input channel that is plotted in Fig. \ref{cond}(a). The conductance of the system is calculated using the Landauer formula
\begin{equation}
G=\frac{e^2}{h}\sum_i^N T_i,
\label{land}
\end{equation}
where the transmission of each $i$'th channel transport mode is calculated from the reflection probability $T_i=N-\sum_j^N R_{i\rightarrow j}$ -- the sum goes over $N$ modes propagating in the $-y$ direction.

For determination of the transport probability we first calculate the available transport modes in the input channel. Since ($\left[ -i\hbar\partial/\partial y, H \right]=0$) the wave vector $k$ is a good quantum
number and the spinor of the electron in the input channel can be written in a separated form
\begin{equation}
\psi^k(x,y)=e^{iky}\left(
\begin{array}{c} \psi_{\uparrow}^k(x)\\
\psi_{\downarrow}^k(x)
\end{array}\right).
\end{equation}
We use this form of the spinor for calculation of the $N$ Fermi wave vectors in the input channel ($y=0$) for each of the transport modes at a given Fermi energy $E_F$. We repeat the procedure for the output channel ($y=4000$ nm) that is 1368 nm wide -- where for $E_F=3$ meV there are 52 subbands [see Fig. \ref{cond}(b) for the dispersion relation] -- obtaining $M$ spinors with the corresponding wave vectors.

In order to determine $R_{i\rightarrow j}$ we solve the Schr\"{o}dinger equation $H\Psi = E\Psi$ for the electron incident from  $i$'th subband with wavevector $k_i$.
The boundary conditions for the ends of the computational box $y=0$, $y=4000$ nm are adopted from Ref. \onlinecite{szafran2011} upon generalization
to the SO coupling case.
In the input lead the electron wave function is a superposition of an incoming wave with $k_l$ and $N$ backscattered waves (with negative currents)
\begin{equation}
\Psi(x,y)=c_l \psi^{k_l}(x,y)+\sum_{j=1}^{N} d_j \psi^{-k_j}(x,y).
\end{equation}
Its derivative has the form,
\begin{equation}
\frac{\partial \Psi(x,y)}{\partial y} = ic_lk_l \psi^{k_l}(x,y)-\sum_{j=1}^N i d_j k_j \psi^{-k_j}(x,y).
\end{equation}
We add to the both sides of the above equation $ik_l\Psi(x,y)$ and replace $\frac{\partial \Psi(x,y)}{\partial y}$ by its central finite difference formula obtaining
the boundary condition at the low end of the computational box
\begin{eqnarray}
\Psi(x,y-\Delta y)=\Psi(x,y+\Delta y)+2\Delta y ik_{l}\Psi(x,y)\nonumber \\
-2i\Delta y \left(2 k_l c_l \psi^{k_l}
 +\sum_{j=1}^{N}  (k_{l}-k_j)d_j \psi^{-k_j}\right)
\end{eqnarray}

At the other end of the computational box there are no backscattered waves,
the wave function has the form
\begin{equation}
\Psi(x,y)=\sum_{j=1}^{N} a_j \psi^{k_j}(x,y),
\end{equation}
and its derivative reads,
\begin{equation}
\frac{\partial \Psi(x,y)}{\partial y} = \sum_{j=1}^{N} i a_j k_j \psi^{k_j}(x,y),
\end{equation}
We follow the procedure applied to the wave function in the incoming lead but now we subtract $ik_{l}\Psi(x,y)$ obtaining
the boundary condition at the top of the computational box
\begin{eqnarray}
\Psi(x,y+\Delta y)=\Psi(x,y-\Delta y)+2\Delta y ik_{l}\Psi(x,y) \nonumber\\
+2i\Delta y  \sum_{j=1}^{N}  (k_j-k_{l})a_j \psi^{k_j}(x,y).
\end{eqnarray}
The expressions for $\Psi(x,y-\Delta y)$ and $\Psi(x,y+\Delta y)$ are introduced to the linear system of equations that
is given by the finite difference form of the Schr\"{o}dinger equation.

We use a finite computational box in our calculations. In order to simulate an infinite semi-plane at the output from QPC we
introduce transparent boundary conditions for the electron waves. For that purpose at the $x$ edges of the computational box we assume the following boundary conditions for $y>1600$ nm [see the red lines in Fig. 1],
\begin{equation}
\Psi(x\pm\Delta x,y)=\Psi(x,y)\exp[ik_{b}\Delta x],
\end{equation}
where ($-$) is for the left boundary and ($+$) for the right one.
The value of the wave vector $k_{b}$ was set to remove the scattering from the edges.
We found that the scattering is nearly removed for $k_b$ which corresponds to a unique wave vector that
appears at the considered Fermi energy $E_F=3$ meV for a thin channel of $W_b=80$ nm.
These boundary conditions allow the electron to flow out freely through the left and right edges of the computational box after QPC leaving the region where the system is scanned by the probe unperturbed -- see the wave function plotted in Fig. \ref{potbmp} with the color map.

We discretize the Hamiltonian Eq. (\ref{ham}) [taking $\Delta x= \Delta y=8$ nm]\cite{disc} and solve the resulting system of linear equations using LU method for sparse matrices.\cite{slu}
The scattering amplitudes $a_j,c_j,d_j$ are determined in a self-consistent manner,\cite{szafran2011} with an initial guess $a_{j}=c_l\delta_{jl}$ and $d_j=0$.
The solution to the linear system of equation provides us with the scattering spinor wave function $\Psi(x,y)$ for a given energy $E_F$.
Before the convergence is reached this wave function contains contributions of all the subbands at a given energy, and the asymptotic form
of Eq. (6) and (9) is only obtained at the self-consistence conditions.
We extract then new values of the
scattering amplitudes by projection of wave function $\Psi(x,y)$ on the eigenmodes $\psi^{k'}$ given by Eq. (5),
for {\it all} transport subbands at the Fermi level.
In particular in the input lead we take
\begin{equation}
\sum_{j=1}^N c_j \langle \psi^{k'} | \psi^{k_j}\rangle+ \sum_{j=1}^N d_j \langle \psi^{k'} | \psi^{-k_j}\rangle=\langle \psi^{k'} | \Psi\rangle.
\label{matr}
\end{equation}
The scalar products in Eq. (\ref{matr}) are evaluated by integration that is performed on 160 nm strips at the beginning and at the end of the channel [see the crossed regions in Fig. \ref{potbmp}]. The new values of the scattering
amplitudes are used for the subsequent iteration of the solution of the Schr\"{o}dinger equation and the procedure is repeated until the convergence is reached, i.e. in the input channel there is only one incoming wave corresponding to $k_l$ and the amplitudes of backscattered waves at the output channel vanish.

Finally, the backscattering probabilities for the Landauer formula are calculated from the probability currents for given wave vectors,
\begin{equation}
R_{i\rightarrow j}=\left| \frac{d_{j}}{c_i}\right|^2\cdot\frac{j_{j}}{j_{i}},
\end{equation}
where the current flux is
\begin{equation}
j_i=\int|\psi^i(x)|^2\frac{\hbar k}{m^*}+\frac{\alpha}{\hbar}\left[\psi_\uparrow^{*i}(x)\psi_\downarrow^i(x)+\psi_\downarrow^{*i}(x)\psi_\uparrow^i(x)\right]dx.
\end{equation}

For numerical calculations we adopt parameters for for $\mathrm{In}_{0.5}\mathrm{Ga}_{0.5}\mathrm{As}$, i.e.  $m^*=0.0465m_0$ and Rashba SO coupling constat that is comparable to the experimentally measured\cite{park} value i.e. $\alpha=11.44$ meVnm (unless stated otherwise) which stems ($\alpha=\alpha_{3D}F_z|e|$) from the material constant\cite{alpha} $\alpha_{3D}=0.572\;\mathrm{nm}^2$ and electric field in the growth direction $F_z=200$ kV/cm. The computational box consist of $171\times501\times2$ points. The present computational scheme gives a direct insight into the solution of the Schr\"odinger equation and the obtained transport properties of the system are equivalent to the ones obtained in Greens function approach.\cite{green}
\section{Results}

\subsection{Conductance of an unperturbed system}

\begin{figure}[!h]
\epsfxsize=80mm
                \epsfbox[35 97 571 740] {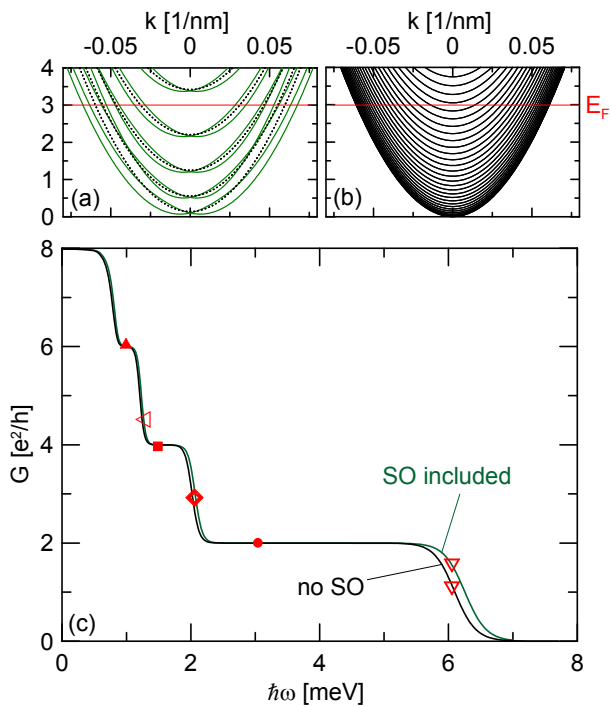}
                 \caption{(color online) (a) Dispersion relation in the input channel without (black dotted curves) and with (green curves) SO interaction included. (b) Dispersion relation in the output channel in the absence of SO coupling. (c) Conductance calculated in the absence (black curve) or in the presence (green curve) of SO coupling. The symbols mark the points in the conductance for which the SGM maps are calculated.}
 \label{cond}
\end{figure}

Let us start with the conductance obtained in the absence of the scanning probe.
The narrowing introduced to the channel by the QPC limits the number of the conducting modes. In Fig. \ref{cond}(c) with the black curve we plotted the conductance versus the QPC potential for $\alpha=0$.
In the absence of the QPC potential -- for $\hbar \omega =0$ [see Eq. (2)] -- we obtain no backscattering. As the QPC is introduced to the system ($\hbar \omega \neq 0$)
the conductance is reduced and the characteristic $G$ plateaux appear. The conductance obtained in the presence of SO interaction is plotted by the green curve in Fig. \ref{cond}(c). We observe that SO coupling does not induce any qualitative changes in the dependence of $G$ on QPC potential and the quantitative modification is also very weak. The green curve is only shifted on the two last steps to higher energies as compared to the dependence obtained in the absence of SO interaction. The similar shape of the $G$ curves obtained with and without SO interaction is a result of the Kramers degeneracy that is preserved in the presence of SO interaction -- despite splitting of the spin modes for each value of Fermi energy there is an even number of conducting modes [compare the green and black dotted curves in Fig. \ref{cond}(a)] which results in quantization of the conductance in $2e^2/h$ steps.


\subsection{Branched electron flow on the conductance plateaux}
\begin{figure*}
\epsfxsize=150mm
                \epsfbox[18 251 583 583] {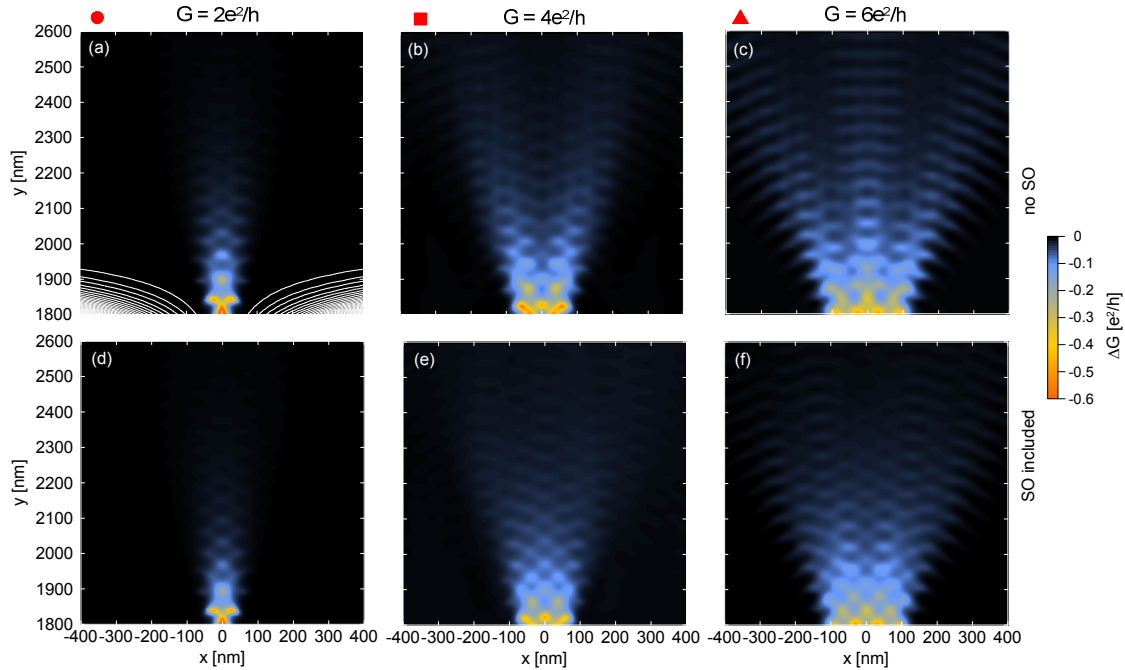}
                 \caption{(color online) Maps of the conductance changes $\Delta G$ as a function of the SGM tip position for the QPC tuned to three plateaux -- the symbols correspond to the ones in Fig. \ref{cond}. The upper (bottom) row corresponds to the results obtained without (with) SO interaction.}
 \label{condmaps}
\end{figure*}

Let us now discuss the maps of the $\Delta G$ -- the difference between the conductance obtained in the presence of the scanning gate tip potential and the unperturbed result -- for the QPC tuned to conductance plateaux. In Figs. \ref{condmaps}(a-c) we present maps for $\hbar\omega=3$ meV, $\hbar\omega=1.6$ meV, and $\hbar\omega=1$ meV -- which correspond to $\bullet,\blacksquare,\blacktriangle$ symbols in Fig. \ref{cond}(c) respectively.
In the absence of SO interaction a clear signature of the angular dependence of the electron flow on the conductance of QPC emerges in the maps as observed in the experiment \cite{topinka2000} performed for GaAs. The number of paths -- a single one [Fig. \ref{condmaps}(a)], two [Fig. \ref{condmaps}(b)], and three [Fig. \ref{condmaps}(c)] corresponds to the number of quantized spin-degenerate orbital modes in the QPC that conduct.
Moreover a strong checkerboard pattern is present\cite{comment1} in the results. It stems from the interference between the waves reflected back by the tip and their reflection from the QPC gates.\cite{jura}

The maps obtained in the presence of SO interaction are displayed in Fig. \ref{condmaps}(d-f). For the QPC tuned to $G=2\;e^2/h$ -- Fig. \ref{condmaps}(d) -- the character of the measured flow does not change. On the other hand there is a significant modification of the maps obtained for more open QPC, namely for $G=4\;e^2/h$ and $G=6\;e^2/h$. Comparing the maps obtained without SO interaction and with SO coupling included we observe that SO interaction results in smearing out the pattern with double and triple paths.

\begin{figure}[ht!]
\epsfxsize=85mm
                \epsfbox[22 153 590 670] {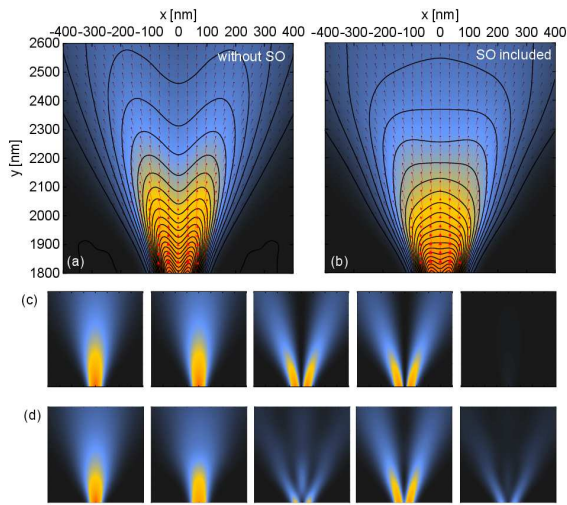}
                 \caption{(color online) (a,b) Color maps show the probability density for the transmitted electron. The arrows present the probability current and the contours present amplitude of the probability current. Results obtained for the plateau $G=4\;e^2/h$ for (a) $\alpha=0$ and for (b) $\alpha=11.44$ meVnm. (c) Probability densities obtained for the transport from the first five modes of the input channel without SO coupling. (d) Same as (c) but for SO interaction included.}
 \label{curtot}
\end{figure}

In order to inspect closer the above finding we focus on $G=4\;e^2/h$.
In Fig. \ref{curtot}(a,b) we plot with the color maps the probability densities, with the arrows the current and with the isolines the amplitude of the current.
Comparing Fig. \ref{condmaps}(b) and Fig. \ref{curtot}(a)
we notice that with the exception of the checkerboard pattern
that is due to the interference introduced by the tip, the SGM conductance maps are well correlated to the current amplitude and probability density of the transmitted electron.
This is also the case when the SO interaction is present as plotted in Figs. \ref{condmaps}(e) and \ref{curtot}(b).

In the map obtained without SO interaction Fig. \ref{curtot}(a) we observe two maxima whereas in the case with SO interaction present depicted in Fig. \ref{curtot}(b) the probability density and the current is approximately constant along the $x$-axis. There is only a slight oscillation present, i.e. the isolines at the bottom of the plot are concave and on the top they are convex. Figs. \ref{curtot}(c) and (d) present the probability densities obtained for the transport from subsequent modes of the input channel obtained without and with SO coupling respectively. For SO interaction included near the QPC all but two firs densities have double maxima character. This results in the concave isolines of the sum of the densities in Fig. \ref{curtot}(b). On the other hand away from QPC we observe an increased share of the nonzero density along the $x=0$ axis away from the QPC as compared to the case of absent SO coupling [see the fifth map in Fig. \ref{curtot}(d)]. As the result in the sum of the densities [see Fig. \ref{curtot}(b)] away from the QPC where the double maxima are spread to the sides a considerable amount of density remain in the center - resulting in the convex isolines.

\begin{figure}[ht!]
\epsfxsize=85mm
                \epsfbox[26 168 570 670] {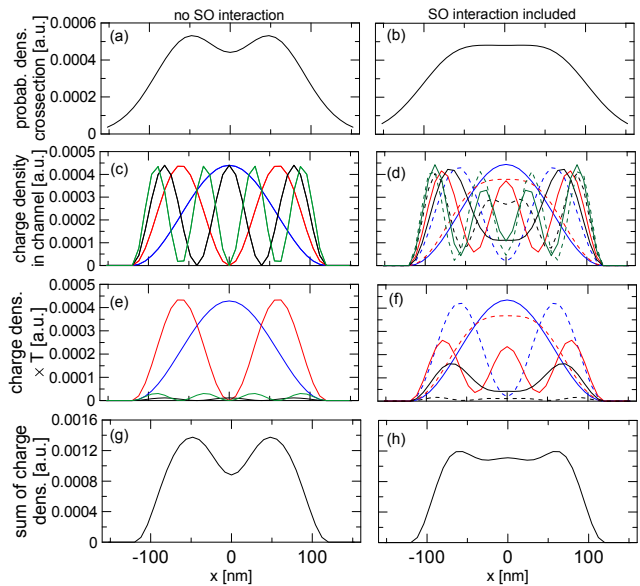}
                 \caption{(color online) (a,b) Cross sections of probability densities above QPC for $y=1896$ nm. (c,d) Charge density corresponding to the eigenmodes of the input channel. (e,f) Densities from (c,d) multiplied by the corresponding transfer probability. (g,h) Sum of the densities from (e,f) correspondingly.}
 \label{denscross}
\end{figure}

To explain the changes in the angular dependence to the densities introduced by SO coupling let us inspect the cross sections of the results obtained for $y=1896$ nm that are presented in Fig. \ref{denscross}(a) in the absence and in Fig. \ref{denscross}(b) in the presence of SO interaction. In Fig. \ref{denscross}(c) we plotted the charge densities of the modes that enter the QPC across the input channel. Each curve corresponds to the density of a spin-degenerate mode. For the QPC tuned to $G=4\;e^2/h$ the transfer probabilities of the subsequent modes are $T_1=T_2=0.97,\;T_3=T_4=0.99,\;T_5=T_6=0.03$, and $T_7=T_8=0$.
 Fig. \ref{denscross}(e) presents the probability densities multiplied by the corresponding transfer  probabilities. We observe that there are two main modes conducting -- the one with a single and two maxima. Fig. \ref{denscross}(g) presents the sum of charge densities from Fig. \ref{denscross}(e). We find that the latter resembles the two-maxima character observed previously at the cross section of the charge density obtained after QPC in Fig. \ref{denscross}(a) so it is the transmission through this two modes that results in the two paths observed in Fig. \ref{condmaps}(b) and Fig. \ref{curtot}(a)

When SO interaction is included the eigenmodes in the input channel are mixed and the spin degeneracy is lifted. Now each mode is split into two and we marked their charge densities with the solid and dashed curves in Fig. \ref{denscross}(d) where the colors of the curves correspond to the ones from Fig. \ref{denscross}(c). The transfer probabilities of these modes are $T_1=0.98, T_2=0.98, T_3=0.63, T_4=0.97, T_5=0.38, T_6=0.04, T_7=0.02, T_8=0$.

\begin{figure}[ht!]
\epsfxsize=90mm
                \epsfbox[24 283 584 558] {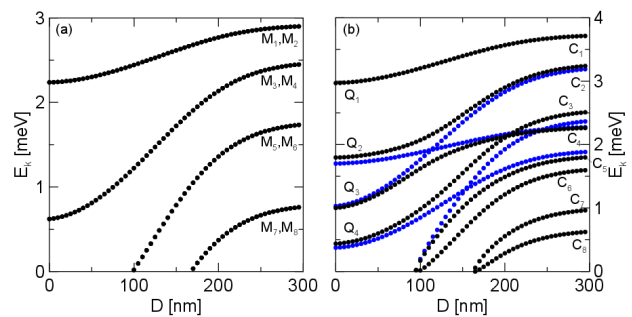}
                 \caption{(color online) Kinetic energy of the modes of channel with transverse potential corresponding to the distance $D$ from the center of the QPC for $E_f=3$ meV. (a) Case without SO interaction. (b) SO interaction included. Black dots corresponds to the case with full SO Hamiltonian included and blue dots corresponds to the case of neglected last term of Eq. (\ref{ham}.)}
 \label{cross}
\end{figure}

The obtained transmission probabilities are no longer approximately binary as the ones obtained without SO interaction. Below we provide a reasoning that explains this observation. Fig. \ref{cross}(a) presents kinetic energies [calculated as $\hbar^2 k^2/(2m^*)$] of the eigenmodes of the channel with the transverse potential that corresponds to the $D$ distance from the QPC center ($y=1600$ nm). Without SO interaction far from the QPC in the input lead there are four spin degenerate modes [$M_1$ to $M_8$ in Fig.\ref{cross}(a)]. As the channel gets narrower a part of the kinetic energy of progressive motion in each mode is converted into energy of the lateral localization within the constriction, which can be treated as an effective potential energy. For the modes with higher lateral excitation the energy drops rapidly. As a result for $D=0$ -- in the middle of the QPC -- there are only two spin-degenerate modes with nonzero kinetic energy and they correspond to the modes of input channel ($M_1,M_2,M_3,M_4$) with the transfer probability close to unity (with $T_1=T_2=0.97,\;T_3=T_4=0.99$). The two other spin-degenerate modes are reflected by the QPC.

For the SO interaction included for $D=300$ nm there are 8 modes with different kinetic energies [see the black dots in Fig. \ref{cross}(b)]. SO interaction lifts the spin degeneracy, but in a channel of finite width the spins are not well defined. The spin mixing is due to $(-\alpha\sigma_y k_x)$ term in the Hamiltonian Eq. (\ref{ham}). With the blue dots in Fig. \ref{cross}(b) we present kinetic energies of four modes obtained when $(-\alpha\sigma_y k_x)$ term is neglected and the spins are well defined in the $x$-direction. We observe that as the $D$ is decreased the energy levels corresponding to the modes with opposite spins cross. For the calculation with full Hamiltonian of Eq. (\ref{ham}) the crossings are replaced by level repulsions due to spin mixing.

Let us now consider electron propagating in the fifth mode [marked with $C_5$ in Fig. \ref{cross}(b)] in the input channel. As $D$ decreases the curve depicted with the black symbols anticrosses with the one corresponding to the third mode $C_3$ at $D=150$ nm. The probability of the transition through the avoided crossing (i.e. mode $C_5$ passes to $Q_4$) depends on the degree of the adiabacity of the transition and the avoided crossing width as predicted by Landau-Zener theory.\cite{lz} For fully adiabatic (diabatic) transition the probability of the transfer from the $C_5$ mode to the $Q_4$ mode is 0 (1). On the other hand only the transition of the $C_5$ mode to $Q_4$ provides that the electron preserves nonzero kinetic energy in the middle of the QPC -- allowing for the transmission through the constriction. Therefore the transfer probability $T_5$ should depend on the transition probability through the anticrossing that in turn is controlled by the rapidness of the energy level changes. The latter is dependent on the spatial span of the QPC\cite{meto} -- that in this case is controlled by $w$. For the value of $w$ considered here ($w=100$ nm) the transfer probability is is $T_4=0.38$. Conversely the transfer probability of the other mode -- $C_3$ -- whose energy participates in the anticrossing is $T_3\simeq1-T_4=0.63$. We performed calculations for different widths of QPC and present the obtained transfer probabilities of the subsequent modes in the Table \ref{tab}. When the QPC length is increased from $50$ nm to $200$ nm -- the case of more adiabatic transition -- the transfer probability of the $C_5$ mode drops from $0.45$ to $0.2$ -- accordingly with the predictions of Landau-Zener theory. On the other hand when the $w$ is changed from $50$ nm to $200$ nm the transfer probability of the $C_3$ rises -- again as expected from the reasoning given above. On the other hand the transfer probabilities of the modes whose kinetic energies either do not go through anticrossings ($C_1$) or go through an anticrossing between the modes that have nonzero kinetic energy in the QPC ($C_2,C_4$) do not change in a significant degree as can be inspected in Table \ref{tab}.

We conclude that the particular values of the transfer probability obtained in the presence of SO interaction result from the transformation of the kinetic energies between the modes that are made possible due to spin mixing and appears as the electron passes the constriction.

\begin{table}
\begin{tabular}{|c|c|c|c|c|c|c|}
  \hline
  QPC length & $T_1$ & $T_2$ & $T_3$ & $T_4$ & $T_5$ & $T_6$\\
  \hline
  50 nm & $0.93$ & $0.9$ & $0.5$ & $0.91$ & $0.45$ & $0.13$ \\
  \hline
  100 nm & $0.98$ & $0.98$ & $0.63$ & $0.97$ & $0.38$ & $0.04$ \\
  \hline
  200 nm & $0.99$ & $0.99$ & $0.81$ & $0.99$ & $0.2$ & $0$ \\
  \hline

\end{tabular}
\caption{Transfer probabilities of the first six channel eigenmodes for different widths of the QPC}
\label{tab}
\end{table}

Let us now go back to the charge densities in the input channel. The particular values of the transmission coefficients result in the amplification of different channel modes as compared to the case without SO interaction -- observe the Fig. \ref{denscross}(f) where we presents the Fermi electron densities in the input channel multiplied by the corresponding transfer probabilities.
We observe that the contribution of the modes with maximum of the charge density in $x=0$ is increased as compared to the case without SO interaction of Fig. \ref{denscross}(e). The sum of the charge densities is presented in Fig. \ref{denscross}(h) and it corresponds to the cross section of the charge density obtained from the solution of the transport problem plotted in Fig. \ref{denscross}(b).
The central local minimum present without SO coupling is replaced by a shallow local maximum.
We conclude that the changes in the conductance maps -- the vanishing of the distinct branches as observed without SO interaction is a result of conduction through the QPC of the modes that posses mixed orbital character in the presence of the SO coupling.

\subsection{Interference fringes on conductance steps}
\begin{figure*}[ht!]
\epsfxsize=150mm
                \epsfbox[19 253 584 584] {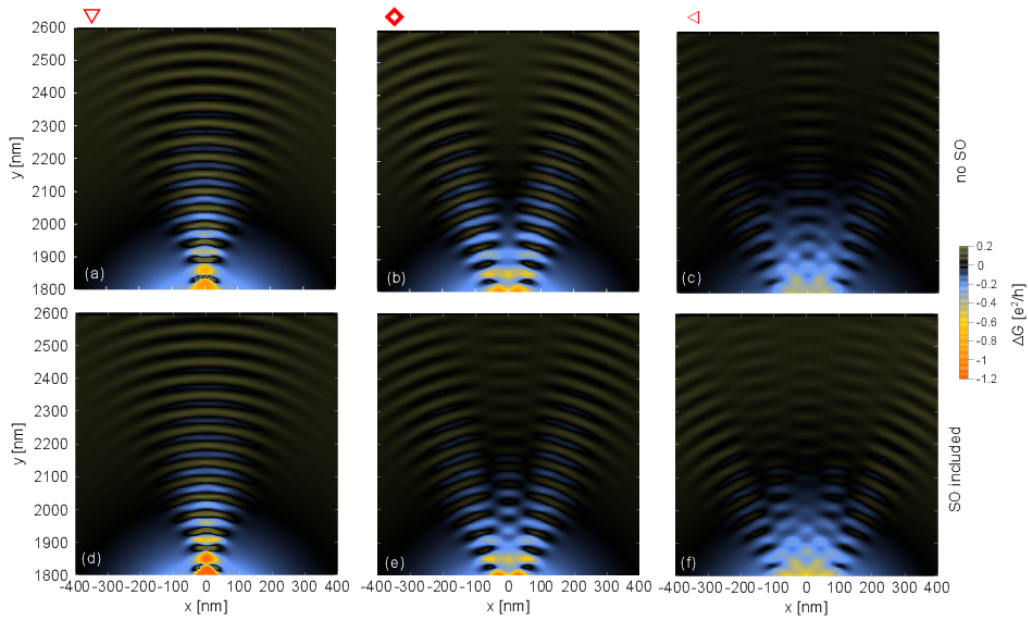}
                 \caption{(color online) Maps of conductance changes obtained with the scanning gate tip potential for the QPC tuned to three conductance steps -- the symbols correspond to the ones in Fig. \ref{cond}. The upper (bottom) row corresponds to the results obtained without (with) SO interaction.}
 \label{condmapsstep}
\end{figure*}
The characteristic feature of the conductance maps obtained in the SGM experiments\cite{topinka2000} on QPC is that the interference fringes are separated by the half of the Fermi wavelength due to interference between the waves reflected by the QPC itself and the tip.\cite{jura}
 These conductance oscillations appear due to interference between the wave function flowing from the constriction and backscattered from the tip. These oscillations in the experiments \cite{topinka2000} are
treated as a signature of the coherent transport.
 The oscillations are most pronounced at the conductance steps -- where the checkerboard pattern -- characteristic to $G$ plateaux are replaced by radial features.\cite{jura}

Figure \ref{condmapsstep} presents the maps of conductance changes $\Delta G$ for the QPC tuned to the conductance steps for $\hbar\omega=6.1$ meV, $\hbar\omega=2$ meV and $\hbar\omega=1.25$ meV that are marked in Fig. \ref{cond} with $\triangledown$, $\lozenge$ and $\vartriangleleft$ respectively.
The results of Fig. \ref{condmapsstep}(a) reproduce the radial fringes obtained in the experimental SGM maps of Ref. \onlinecite{jura}.
We notice that the calculated changes of the conductance are two times larger as compared with the maps obtained at the $G$ plateaux [see Fig. \ref{condmaps}]. Moreover, the tip now induces also an increase in conductance far from the QPC -- the positive changes are denoted with the brown colors in the maps of Figs. \ref{condmapsstep}.

For QPC tuned to the conductance step below the last plateaux $G<2\;e^2/h$ there are no quantitative changes between the results obtained without SO coupling Fig. \ref{condmapsstep}(a) and with SO interaction present Fig. \ref{condmapsstep}(d). On the other hand the difference is visible comparing the maps for the QPC tuned to the conductance step between $G=2\;e^2/h$ and $G=4\;e^2/h$ plotted in Figs. \ref{condmapsstep}(b,e) and for the QPC tuned to the conductance step between $G=4\;e^2/h$ and $G=6\;e^2/h$ displayed in Figs. \ref{condmapsstep}(c,f). Here, we observe that the $\Delta G$ maps bear the signature of mode mixing as there is nonzero flow present along the symmetry axis of the QPC in Fig. \ref{condmapsstep}(e) and five electron flow paths are present in Fig. \ref{condmapsstep}(f).

\begin{figure}[ht!]
\epsfxsize=85mm
                \epsfbox[28 317 562 525] {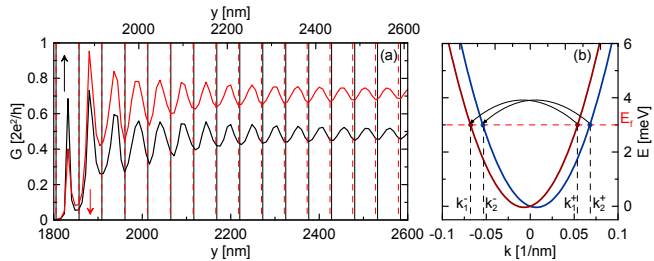}
                 \caption{(color online) (a) Conductance as a function of scanning gate position along $x=0$ axis obtained for the QPC tuned to the last conductance step for $\hbar\omega=6.1$ meV without (black curve) and with SO interaction (red curve). The vertical solid and dashed lines depict $l$ and $l_{SO}$ respectively [see text]. (b) Dispersion relation of the 2DEG in the presence of SO interaction. Colors of the curves represent spin polarization antiparallel (blue) and parallel (red) to the $x$-direction. The curved arrows presents the transitions between the modes that occur during the backscattering from the tip.}
 \label{osc}
\end{figure}

The interference fringes that are present in the maps are separated by the half of Fermi wavelength ($l$) for unconfined 2DEG. The energy reads
\begin{equation}
E=\frac{\hbar^2 k_F^2}{2m^*},
\end{equation}
then $l=\lambda_F=\pi/k_F$, $k_F=\sqrt{2m^*E_F/\hbar^2}$ and we obtain
\begin{equation}
l=\frac{\pi \hbar}{\sqrt{2m^*E_F}},
\end{equation}
which for $E_F=3$ meV gives $l=51.92$ nm. In Fig. \ref{osc} with the black curve we show the cross section of the conductance corresponding to map of Fig. \ref{condmapsstep}(a) obtained for $x=0$. We mark the oscillation period $l$ with the solid black vertical lines.

In the presence of SO interaction the single parabola in the dispersion relation is split into two, each one corresponding to the opposite spin polarization. We depict the dispersion relation in Fig. \ref{osc}(b) with the blue and red curves. In the present case for the electron propagating along the $y$-direction (with $k$ being its wavevector) in presence of the  Rashba coupling the spin is polarized in the $x$-direction and the dispersion relation consists of two branches [see Fig.\ref{osc}(b)],
\begin{equation}
\begin{split}
E_-=\frac{\hbar^2 k^2}{2m^*}-\alpha k,\\
E_+=\frac{\hbar^2 k^2}{2m^*}+\alpha k,
\end{split}
\end{equation}
which gives four possible values of the wavevector:
\begin{equation}
\begin{split}
k^{\pm}_1=\left(\pm \sqrt{\alpha^2{m^*}^2+2E_F m^*\hbar^2} -\alpha m^*\right)/\hbar^2,\\
k^{\pm}_2=\left(\pm \sqrt{\alpha^2{m^*}^2+2E_F m^*\hbar^2}+\alpha m^*\right)/\hbar^2.
\end{split}
\end{equation}
There are two positive wave vectors: $k^{+}_1=(\sqrt{\alpha^2{m^*}^2+2E m^*\hbar^2}-\alpha m^*)/\hbar^2$ and $k^{+}_2=(\sqrt{\alpha^2{m^*}^2+2E m^*\hbar^2}+\alpha m^*)/\hbar^2$. Therefore one might expect that the two close frequencies of the oscillations should disturb the interference fringes by forming a beating pattern --  as it is the case for Zeeman splitting.\cite{klesh} In Fig. \ref{osc}(a) with the red curve we present the cross section of the conductance corresponding to the map of Fig. \ref{condmapsstep}(d). Nevertheless, we find an oscillation with a single frequency as in the case of absent SO coupling. We find that the backscattering by the tip does not induce significant spin flips. As a result when the electron is backscattered by the tip
its wave vector changes not only sign but also the absolute value -- see the transitions marked with the arrows in Fig. \ref{osc}(b). The fringes observed in the SGM map are due to formation of standing waves between the tip and the QPC. Since on one way the electron
travels with wave vector $k^+_2$ and the other with $k^-_2$ the entire phase shift on the back and forth travel can be accompanied to the average wave vector,
\begin{equation}
k_a=(k^{+}_2-k^{-}_2)/2=\sqrt{\frac{2E_F m^*}{\hbar^2}+\frac{\alpha^2{m^*}^2}{\hbar^4}}.
\label{shift}
\end{equation}
 The analogical proces takes place for the electron propagating with $k^+_1$ and again we obtain the same average wavevector $k_a=(k^{+}_1-k^{-}_1)/2$. Therefore the transmission calculated for each mode exhibits exactly the same oscillation pattern and finally the conductance oscillations have period that corresponds to the value of $l_{SO}=\pi/k_a$, i.e. half of the wavelength of the average $k$. In the present case $l_{SO}=51.58$ nm and we mark this period with the red vertical lines in Fig. \ref{osc}(a). The obtained period is similar to the one found without SO coupling due to small value of the last term in the square root in Eq. \ref{shift} as compared to the first term. We conclude that in contrast to the changes of the angular flow the SO interaction does not significantly alter the interference fringes. The obtained oscillations have period that is similar to the one obtained in the absence of SO coupling and this fact corresponds to the reflection of the electron waves without spin flips.

\subsubsection{Conductance oscillation in the presence of Zeeman effect}
\begin{figure}[ht!]
\epsfxsize=85mm
                \epsfbox[11 160 570 680] {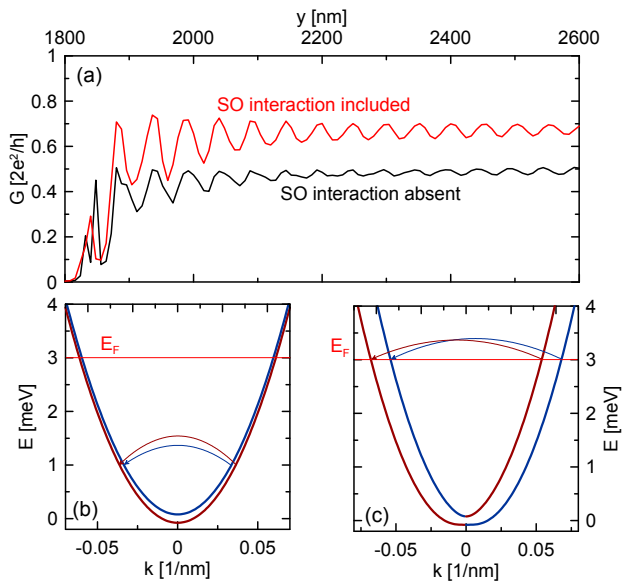}
                 \caption{(color online) (a) Conductance as a function of scanning gate position along $x=0$ axis obtained for the QPC tuned to the last conductance step for $\hbar\omega=6.1$ meV in the presence of Zeeman term for B=0.3 T. Black (red) curve corresponds to the case $\alpha=0$ ($\alpha=11.44$ meVnm). (b) Dispersion relation for Zeeman effect present with $\alpha$=0. The colors of the curves depict positive and negative spin polarization in the $z$-direction. (c) Dispersion relation for Zeeman effect present with $\alpha$=11.44 meVnm. The colors of the curves depicts the sign of the $\langle s_x\rangle$.}
 \label{zeeman}
\end{figure}

Reference \onlinecite{klesh} reported that due to Zeeman spin splitting the radial fringes in the conductance changes undergo a beating pattern. When SO interaction is absent the Zeeman interaction splits the subbands of the dispersion relation of unconfined 2DEG -- see Fig. \ref{zeeman}(b). Reflection of the wave propagating along the channel back to the QPC by the tip does not change the electron spin [observe the arrows in Fig. \ref{zeeman}(b)] and in the results both the waves have the same absolute value of the momentum (that depends on the spin polarization). Therefore transport in each spin mode will result in an oscillation pattern with a different period and hence the total conductance undergoes the beating due to superposition of transfer probabilities oscillating with different periods. The beating pattern is reproduced in our calculation when we add the Zeeman term $H_z = -1/2 g \mu_0 B \sigma_z$ to the Hamiltonian Eq. (\ref{ham}) for $B=0.3$ T and $\alpha=0$ [observe black curve in Fig. \ref{zeeman}(a)].

On the other hand for nonzero $\alpha$ the spin is no longer well defined. We find that in the present case the mean value of the spin operator in the $x$-direction is of order of magnitude larger than the mean value of $\langle s_z \rangle$ (we obtain for the mode with larger wavevector values $\langle s_x \rangle = 0.995\;\hbar/2$ and $\langle s_z \rangle = 0.099\;\hbar/2$). The dispersion relation is presented in Fig. \ref{zeeman}(c) where the color of the curves depict sign of the $\langle s_x \rangle$. In this case the reflection by the tip preserve the value $\langle s_x \rangle$ and results in change of the momentum of the wave propagating from the QPC and the one reflected by the tip -- as described above. This is reflected by vanishing of the beating pattern as can be observed in Fig. \ref{zeeman}(a) where with the red curve we plot the conductance changes in the presence of both Zeeman effect and SO coupling.

\section{Conclusions}
In summary we described the impact of the spin-orbit interaction on scanning gate microscopy of the electron flow from a QPC.
We demonstrated that
that the SGM maps gathered at the conductance plateaux for $G>2\;e^2/h$ loose their distinct angular pattern.
We explained that this is due to conduction through the QPC of the modes that posses mixed orbital character in the presence of SO coupling.
The maps measured at the conductance steps also bear signatures of the mode mixing but the distinct radial fringe pattern is not destroyed by the SO interaction despite the presence of two different Fermi wavelengths. We find that when SO interaction is present the fringes are separated by a length that corresponds to the mean value of the two Fermi wave vectors. This indicates that the backscattering from the tip is a spin preserving process in the presence of SO interaction. We find moreover that SO coupling suppresses the beating pattern appearing in the presence of pure Zeeman effect.

\section*{Acknowledgements}
This work was supported National Science Centre according to decision DEC-2012/05/B/ST3/03290, and by PL-Grid Infrastructure. MPN gratefully acknowledges the support from the Foundation for Polish Science (FNP) under START programme.

\section{Appendix}

\subsection{Impact of the QPC span}
\begin{figure}[ht!]
\epsfxsize=85mm
                \epsfbox[61 204 585 640] {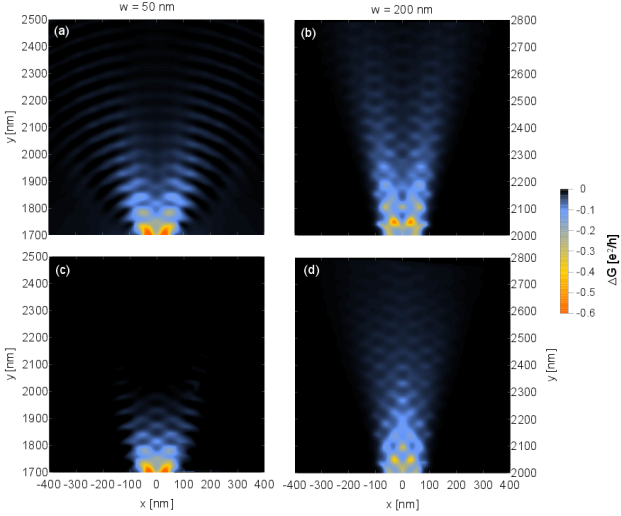}
                 \caption{(color online) Maps of the conductance changes $\Delta G$ as a function of the SGM tip position for the QPC tuned to G = 4 $e^2/h$ obtained for QPC potential of different length. The upper (bottom) row corresponds to the results obtained without (with) SO interaction.}
 \label{length}
\end{figure}

As discussed in the main text the transfer probabilities of the particular modes of the input channel depends on the QPC length ($w$). Here we present the influence of $w$ on the maps of the conductance changes probed by SGM. Fig. \ref{length} presents maps of $\Delta G$ obtained for $w=50$ nm and $w=200$ nm. Comparing the results of Fig. \ref{length}(a)(b) and Fig. \ref{condmaps}(b) we observe that in the absence of SO interaction for shorter QPC the two branches are more open. Also the checkerboard patter gets weaker as the electron flow is now less concentrated.
In the presence of SO interaction the effect of smearing of the angular pattern is present independently of the QPC length as depicted in Figs. \ref{length}(c)(d).

\subsection{Channel width and SO coupling strength}

\begin{figure}[ht!]
\epsfxsize=85mm
                \epsfbox[20 172 587 660] {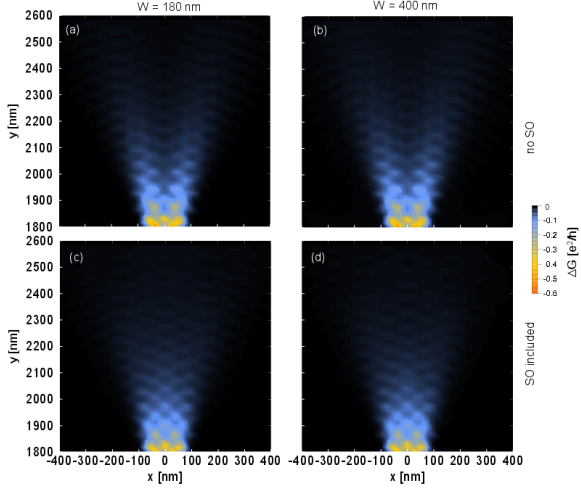}
                 \caption{(color online) Maps of the conductance changes $\Delta G$  for varied width $W$ of the input channel for the QPC tuned to $G=4\;e^2/h$.}
 \label{diffw}
\end{figure}

The width of the input channel $W$ modifies the number of the eigenstates for a given  $E_F$.
In the experimental situation the QPCs usually separate two 2DEG semiplanes so $W$ is very large.
In Fig. \ref{diffw} we show the SGM maps for the QPC tuned to $G=4\;e^2/h$ for two widths of the input channel --  $W=180$ nm and $W=400$ nm. We obtain results that are nearly identical with the ones presented in the main text in Figs. \ref{condmaps}(b,e) for $W=240$ nm.
Therefore, the discussed effects of smearing of the current flow as discussed above is not an effect specific to a particular width of the input channel.

\begin{figure}[ht!]
\epsfxsize=85mm
                \epsfbox[11 277 587 563] {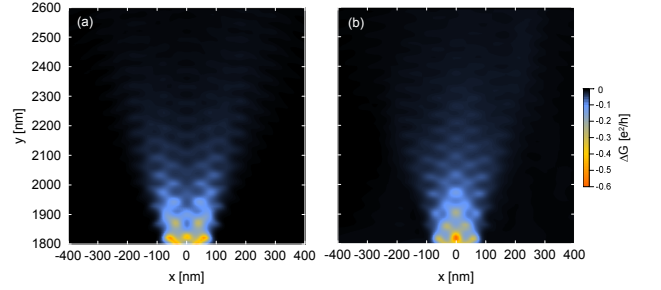}
                 \caption{(color online) Maps of the conductance changes $\Delta G$  for (a) $\alpha=5.72$ meVnm and (b) $\alpha=17.16$ meVnm for the QPC tuned to $G=4\;e^2/h$.}
 \label{difffz}
\end{figure}

Figures \ref{difffz}(a,b) presents the SGM maps for two different strengths of the SO interaction.
Comparing the plots with Fig. \ref{condmaps} (e) for $\alpha=11.44$ meVnm we observe that as the SO interaction gets stronger the electron flow along
the symmetry axis of the QPC is enhanced.


\begin{thebibliography}{00}
\bibitem{disc} Convergence and accuracy of the present results obtained for grid spacing 8 nm was confirmed by comparison with the results obtained on a more dense grid.
\bibitem{comment1} The obtained checkerboard pattern is not a result of finite width of the channel. We have compared present results with the ones obtained for increased width of the channel and the conductance pattern remain unchanged.
\end{thebibliography}
\end{document}